%% file: main.tex
\documentclass[cits]{PoS}

\usepackage[utf8]{inputenc}
\usepackage{amsmath,amssymb,fontenc,enumerate,times,mathptmx,graphicx,float}
\usepackage[backend=bibtex,url=false,doi=false,firstinits=true,hyperref,style=numeric,sorting=none]{biblatex}
\AtEveryBibitem{\clearfield{title}}

\bibliography{ref.bib}

\usepackage[nodisplayskipstretch]{setspace}

\newcommand{\RBRC}{
  RIKEN BNL Research Center,
  Brookhaven National Laboratory,
  Upton, New York 11973,
  USA}

\newcommand{\UCONN}{
  Physics Department,
  University of Connecticut,
  Storrs, Connecticut 06269-3046,
  USA}

\newcommand{\NAGOYA}{
  Department of Physics,
  Nagoya University,
  Nagoya 464-8602,
  Japan}
  
\newcommand{\NISHINA}{
  Nishina Center,
  RIKEN,
  Wako, Saitama 351-0198,
  Japan}

\newcommand{\BNL}{
  Physics Department,
  Brookhaven National Laboratory,
  Upton, New York 11973,
  USA}

\newcommand{\CU}{
  Physics Department,
  Columbia University,
  New York, New York 10027,
  USA}

\title{The connected and leading disconnected diagrams of the hadronic light-by-light contribution to muon $g-2$}

\ShortTitle{Hadronic Light-by-Light in Muon $g-2$}

\author{\speaker{Luchang Jin}\\
        \CU\\
        E-mail: \email{ljin.luchang@gmail.com}}

\author{Thomas Blum\\
{\UCONN}\\
{\RBRC}}

\author{Norman Christ\\
  {\CU}}

\author{Masashi Hayakawa\\
{\NAGOYA}\\
{\NISHINA}}

\author{Taku Izubuchi\\
{\BNL}\\
{\RBRC}}

\author{Chulwoo Jung\\
{\BNL}}

\author{Christoph Lehner\\
{\BNL}}

\abstract{
We report our recent lattice calculation of hadronic light-by-light contribution to muon $g-2$ using our recently developed moment method. The connected diagrams and the leading disconnected diagrams are included. The calculation is performed on a $48^3\times96$ lattice with physical pion mass and 5.5 fm box size. We expect sizable finite volume and finite lattice spacing corrections to the results of these calculations which will be estimated in calculations to be carried out over the next 1-2 years.
}

\FullConference{34th annual International Symposium on Lattice Field Theory\\
		24-30 July 2016\\
		University of Southampton, UK}

\begin{document}

\setlength{\abovedisplayskip}{0.2cm}
\setlength{\belowdisplayskip}{0.2cm}

\input{body.tex}

\setlength\bibitemsep{0.5\itemsep}
\printbibliography{}

\end{document}

%% file: body.tex
\section{Introduction}\vspace{-0.3cm}

\begin{figure}[H]
  \begin{center}
    \resizebox{0.3\columnwidth}{!}{\includegraphics{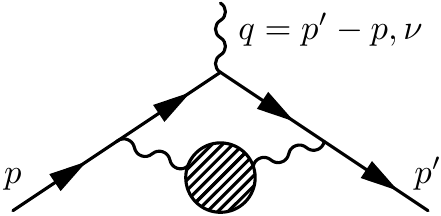}}
    \resizebox{0.3\columnwidth}{!}{\includegraphics{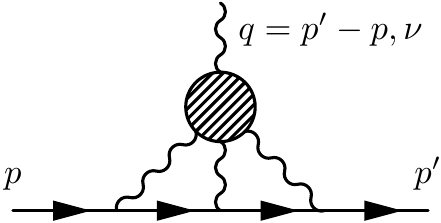}}
  \end{center}
  
\vspace{-0.6cm}
  \caption{\label{fig:hvp-hlbl}(Left) Hadronic vacuum polarization (HVP).
  (Right) Hadronic light-by-light (HLbL).}
\end{figure}

The anomalous magnetic moment of muon, $a_{{\mu}}$, can be measured very
precisely by experiments. The dimensionless number, $a_{{\mu}}$, is
defined through the following relation:
\begin{eqnarray}
  \vec{{\mu}} & = & 2 (1 + a_{{\mu}})  \frac{- e}{2 m_{{\mu}}} 
  \vec{s} 
\end{eqnarray}
where $\vec{{\mu}}$ is the magnetic moment of muon and $\vec{s}$ is its
spin. The past experiment, BNL E821 {\cite{Bennett:2006fi}}, has obtained
$a_{{\mu}}^{\text{exp}} = 11659208.0 (6.3) \times 10^{- 10}$. However,
this value is larger than the standard model prediction, whose uncertainty is
estimated to be around $5.9 \times 10^{- 10}$, by three standard deviations
{\cite{Jegerlehner:2009ry}}. Before one can declare the discovery of physics
beyond the standard model, both the experimental value and the theoretical
prediction need to be improved. Therefore, much more accurate experiments,
Fermilab E989 {\cite{Carey:2009zzb}} and J-PARC E34 {\cite{Mibe:2010zz}}, are
under active preparation. The experimental uncertainties are expected be
reduced to $1 / 4$ of its current value. On the other hand, in order to
improve the accuracy of the standard model prediction, one need to address two
kinds of diagrams shown in Fig. \ref{fig:hvp-hlbl}, namely the hadronic vacuum
polarization (HVP) diagram and the hadronic light-by-light (HLbL) diagram,
which are the major sources of uncertainty in the current theoretical
prediction.\vspace{-0.3cm}

\vspace{-0.4cm}

In this proceeding, we will discuss the lattice calculation of the connected
and leading disconnected hadronic light-by-light amplitude. Previously this
quantity has only been calculated using models {\cite{Prades:2009tw}}.
Attempts using lattice QCD were begun by T. Blum, M. Hayakawa, and T. Izubuchi
more than 6 years ago {\cite{Hayakawa:2005eq,Blum:2014oka}}. We have
improved the methodology dramatically, as described in Refs
{\cite{Jin:2015eua,Blum:2015gfa,Blum:2016lnc}}, which leads to
a reduction in statistical errors by more than an order of magnitude, and
includes the leading disconnected-diagram contribution for the first time.
Since much of the material that was presented at LATTICE 2016 has now appeared
in a recent paper {\cite{Blum:2016lnc}}, this proceeding, after a short
summary of the connected and leading disconnected diagram calculation, is
devoted to an expanded discussion of a topic that was only briefly presented
during the conference: the large separation behaviour of the
four-point-function within the light-by-light diagram.\vspace{-0.3cm}

\section{Lattice calculation setup and results}\vspace{-0.3cm}

\begin{figure}[H]
  \begin{center}
    \resizebox{0.23\columnwidth}{!}{\includegraphics{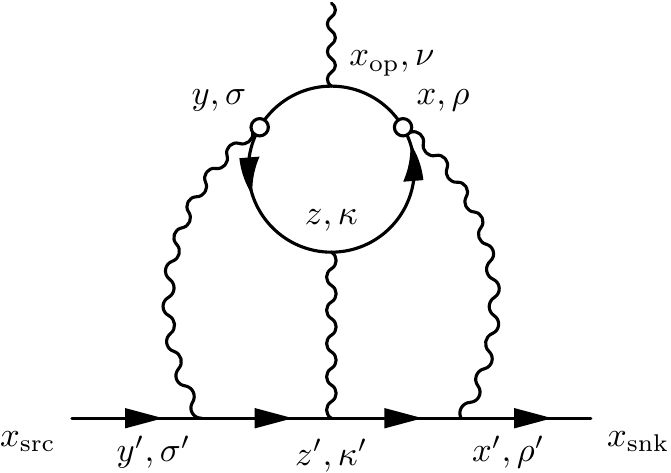}}
    \resizebox{0.23\columnwidth}{!}{\includegraphics{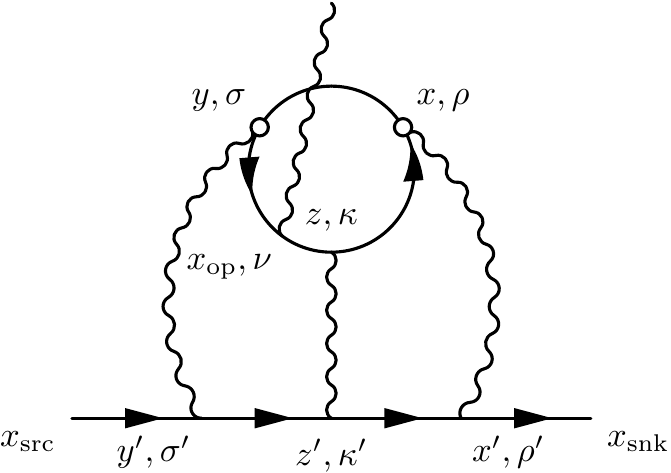}}
    \resizebox{0.23\columnwidth}{!}{\includegraphics{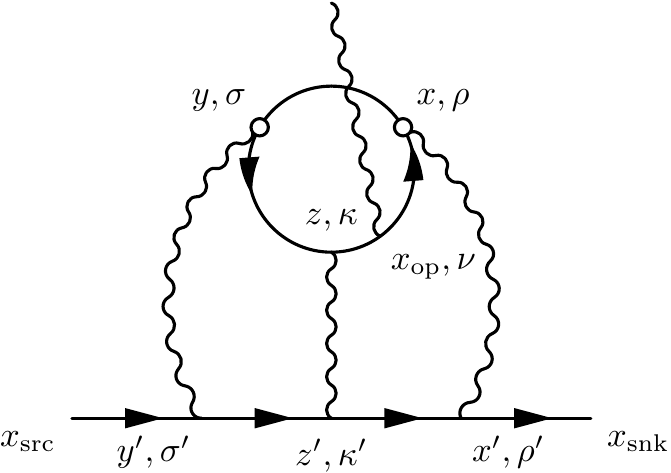}}
    \resizebox{0.23\columnwidth}{!}{\includegraphics{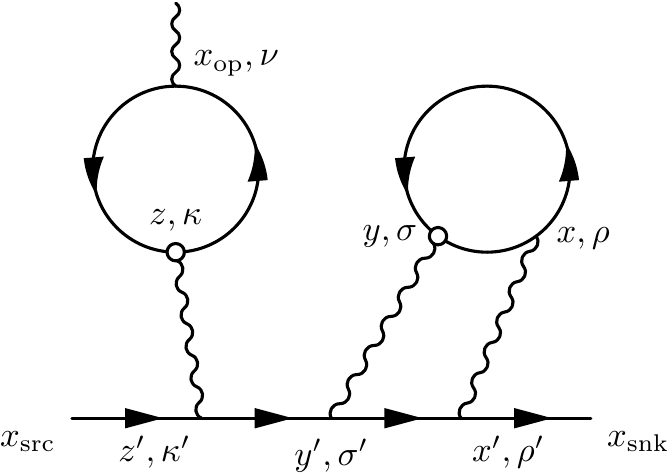}}
  \end{center}
  
\vspace{-0.7cm}
  \caption{\label{fig:c-3}The first three diagrams show the three different
  ways of inserting the external photon when the vertices $x$ and $y$ are
  fixed. The forth diagram is the leading order disconnected diagram, the only
  type of disconnected diagrams that survies in SU(3) limit, other
  disconnected diagrams are not included in the calculation yet. They are
  listed in Ref. {\cite{Blum:2016lnc}}. For each of these four diagrams, there
  are five other possible permutations of the connections between the three
  internal photons and the muon line that are not shown.}
\end{figure}

We start the discussion by repeating our final moment-method formula for
evaluating the connected light-by-light contribution to
$a_{{\mu}}^{\ensuremath{\operatorname{cHLbL}}}$ obtained in Refs
{\cite{Jin:2015eua,Blum:2015gfa,Blum:2016lnc}}.
\begin{eqnarray}
  a_{{\mu}}^{\ensuremath{\operatorname{cHLbL}}}  \frac{(\sigma_{s'
  s})_i}{2 m_{{\mu}}} & = & \sum_{r, \tilde{z}} \mathfrak{Z} \left(
  \frac{r}{2}, - \frac{r}{2}, \tilde{z} \right) \sum_{\tilde{x}_{\text{op}}} 
  \frac{1}{2} \epsilon_{i j k} \left( \tilde{x}_{\text{op}} \right)_j \cdot i
  \bar{u}_{s'} (\vec{0}) \mathcal{F}^C_k \left( \frac{r}{2}, - \frac{r}{2},
  \tilde{z}, \tilde{x}_{\text{op}} \right) u_s (\vec{0}), 
  \label{eq:f2-lbl-moment-short-z}
\end{eqnarray}
where $(\sigma_{s' s})_i = \bar{u}_{s'} (\vec{0}) \Sigma_i u_s (\vec{0})$ are
the conventional Pauli matrices, $\Sigma_k = \frac{1}{4 i} \epsilon_{i j k}
[\gamma_i, \gamma_j]$, and the weight function ``$\mathfrak{Z}$'' is defined
in Ref. {\cite{Jin:2015eua,Blum:2016lnc}}, but could be replaced by
$1$ without hurting correctness. The integration variables are related to the
coordinates in the first three diagrams in Figure \ref{fig:c-3} by the
following equations: $r = x - y$, $\tilde{z} = z - (x + y) / 2$,
$\tilde{x}_{\text{op}} = x_{\text{op}} - (x + y) / 2$. The function
$\mathcal{F}^C_{\nu} \left( x, y, z, x_{\text{op}} \right)$ represent the
Feynman amplitude; an explicit expression is given in Ref
{\cite{Blum:2016lnc}}.\vspace{-0.4cm}

\vspace{-0.5cm}

We perform the sum over $r = x - y$ by sampling a few values on each QCD
configuration and use point source propagators at $x$ and $y$. All other
coordinates are summed over completely on the lattice using the point source
propagators and the sequential source propagators. Since $x$ and $y$ are
connected by quark lines, it can be expected that the major contribution to $g
- 2$ comes from the region where $| r |$ is small. We accommodate this by
using importance sampling. We even perform a complete sum, up to discrete
symmetries, in the region where $| r | \leqslant r_{\mathrm{\max}}$.
$r_{\text{max}}$ is chosen to be $5$ in lattice units in our numerical
simulations.

For the disconnected diagrams in Figure \ref{fig:c-3}, the moment method can
still be applied.
\begin{eqnarray}
  a_{{\mu}}^{\ensuremath{\operatorname{dHLbL}}}  \frac{(\sigma_{s',
  s})_i}{2 m_{{\mu}}} & = & \sum_{r, \tilde{x}}
  \sum_{\tilde{x}_{\text{op}}}  \frac{1}{2} \epsilon_{i, j, k} \left(
  \tilde{x}_{\text{op}} \right)_j \cdot i \bar{u}_{s'} (\vec{0})
  \mathcal{F}^D_k \left( \tilde{x}, 0, r, r + \tilde{x}_{\text{op}} \right)
  u_s (\vec{0}) .  \label{eq:f2-dlbl-moment}
\end{eqnarray}
The integration variables are related to the coordinates in the forth diagram
in Figure \ref{fig:c-3} by the following equations: $r = z - y$, $\tilde{x} =
x - y$, $\tilde{x}_{\text{op}} = x_{\text{op}} - z$. Just like the connected
diagram calculation, the sum over $r$ is performed by randomly sampling the
$(z, y)$ point-pairs. In order to increase statistics, we compute point source
propagators for $1024$ randomly (not uniformly, see Ref {\cite{Blum:2016lnc}})
chosen points and use all possible $1024^2$ combinations as the sample of
point-pairs. The amplitude $\mathcal{F}^D_{\nu} \left( x, y, z, x_{\text{op}}
\right)$ is given by:
\begin{eqnarray}
  \mathcal{F}^D_{\nu} \left( x, y, z, x_{\text{op}} \right) & = & (- i e)^6
  \mathcal{G}_{\rho, \sigma, \kappa} (x, y, z) \left\langle \frac{1}{2}
  \Pi_{\nu, \kappa} \left( x_{\text{op}}, z \right)  \left[ \Pi_{\rho, \sigma}
  (x, y) - \Pi^{\text{avg}}_{\rho, \sigma} (x - y) \right]
  \right\rangle_{\text{QCD}} .  \label{eq:dlbl-amp}
\end{eqnarray}
The function $\mathcal{G}_{\rho, \sigma, \kappa} (x, y, z)$ represents the
photon and muon part of the diagram {\cite{Blum:2016lnc}}. The function
$\Pi_{\rho, \sigma} (x, y)$ is given by the following expression:
\begin{eqnarray}
  \Pi_{\rho, \sigma} (x, y) & = & - \sum_q (e_q / e)^2
  \ensuremath{\operatorname{Tr}} [\gamma_{\rho} S_q (x, y) \gamma_{\sigma} S_q
  (y, x)] . 
\end{eqnarray}
The subtraction term in Eq. (\ref{eq:dlbl-amp}) does not contribute to the
central value {\cite{Blum:2016lnc}}. We only perform this subtraction in our
lattice calculation as a noise reduction technique, and we are free to choose
the subtraction term $\Pi^{\ensuremath{\operatorname{avg}}}$ as long as it is
a pre-determined constant in the ensemble average process. Note that a similar
subtraction would be essential if the moment method is not applied
{\cite{Hayakawa:2015ntr}}.

\begin{figure}[H]
  \begin{center}
    \resizebox{0.4\columnwidth}{!}{\includegraphics{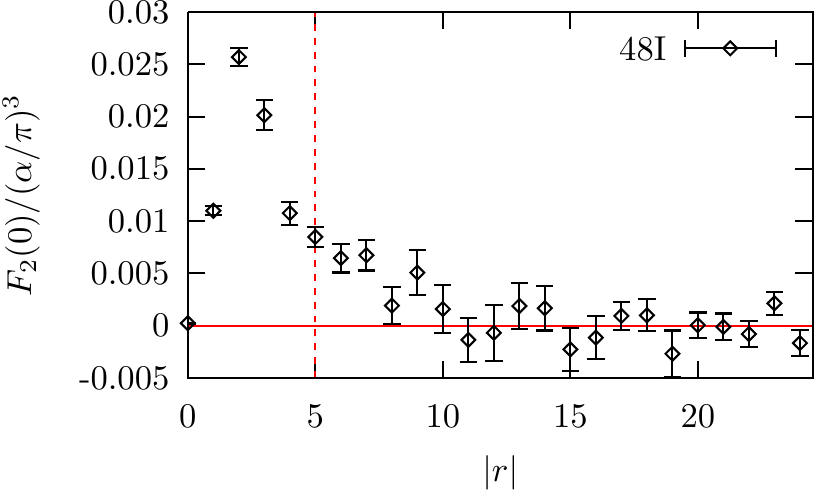}}
    \
    \resizebox{0.4\columnwidth}{!}{\includegraphics{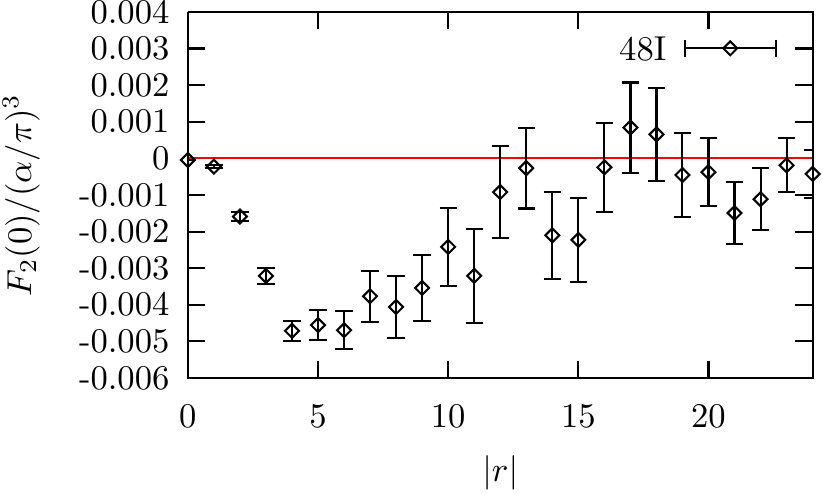}}
  \end{center}
\vspace{-0.7cm}
  \caption{\label{fig:histogram}Histograms of the contribution to $F_2$ from
  different separations $| r |$. The $i$th bin includes contributions from the
  region where $i - 1 < | r | \leqslant i$. The sum of all these bins gives
  the final result for $F_2$. Left: cHLbL contribtion and $r = x - y$. Right:
  dHLbL contribution and $r = z - y$.}
\end{figure}

The calculations were performed on the $48^3 \times 96$ physical-pion-mass
ensemble {\cite{Blum:2014tka}} generated by the RBC and UKQCD collaborations.
We have performed the calculation it on 65 configurations each separated by 20
MD units. Based on our strategy stated above, we sample point-pairs to perform
the sum over $r$. In Figure \ref{fig:histogram}, we plot the histograms based
on those point-pairs.

The sum of the contributions from all separations gives the result:
$a_{{\mu}}^{\ensuremath{\operatorname{cHLbL}}} = (11.60 \pm 0.96) \times
10^{- 10}$, and $a_{{\mu}}^{\ensuremath{\operatorname{dHLbL}}} = (- 6.25
\pm 0.80) \times 10^{- 10}$. Thus
\begin{eqnarray}
  a_{{\mu}}^{\ensuremath{\operatorname{HLbL}}} & = & (5.35 \pm 1.35)
  \times 10^{- 10} .  \label{eq:a-hlbl}
\end{eqnarray}
The errors are statistical only. It might come as a surprise that the
disconnected diagram contribution is quite large. We will show in the next
section that this result is more or less expected.\vspace{-0.3cm}

\vspace{-1.2cm}

\section{Connections between the connected and disconnected
diagrams}\vspace{-0.4cm}

\subsection{The $\pi^0$ exchange diagram}\vspace{-0.2cm}

We know that $\pi^0$ is the lightest particle in QCD, because of the
spontaneous chiral symmetry breaking and Goldstone mechanism. For two points
$x$, $y$ separated by long distance
\begin{eqnarray}
  \langle (\bar{u} \gamma_5 u - \bar{d} \gamma_5 d) (x) (\bar{u} \gamma_5 u -
  \bar{d} \gamma_5 d) (y) \rangle & \sim & e^{- m_{\pi} | x - y |}, \\
  \langle (\bar{u} \gamma_5 u + \bar{d} \gamma_5 d) (x)  (\bar{u} \gamma_5 u +
  \bar{d} \gamma_5 d) (y) \rangle & \sim & e^{- m_{\eta} | x - y |} . 
\end{eqnarray}
Assuming perfect isospin symmetry, the above relations imply:
\begin{eqnarray}
  \langle \bar{u} \gamma_5 u (x) (\bar{u} \gamma_5 u - \bar{d} \gamma_5 d) (y)
  \rangle & \sim & e^{- m_{\pi} | x - y |}, \\
  \langle \bar{u} \gamma_5 u (x) \bar{d} \gamma_5 d (y) \rangle + \frac{1}{2}
  \langle \bar{u} \gamma_5 u (x) (\bar{u} \gamma_5 u - \bar{d} \gamma_5 d) (y)
  \rangle & \sim & e^{- m_{\eta} | x - y |} . 
\end{eqnarray}
This tells us a relationship between the connected diagram and the
disconnected diagram:
\begin{eqnarray}
  \langle \bar{u} \gamma_5 u (x) \bar{d} \gamma_5 d (y) \rangle & = & -
  \frac{1}{2} \langle \bar{u} \gamma_5 u (x) (\bar{u} \gamma_5 u - \bar{d}
  \gamma_5 d) (y) \rangle [1 +\mathcal{O} (e^{- (m_{\eta} - m_{\pi}) | x - y
  |})]  \label{eq:neutral-pion}
\end{eqnarray}
where the left hand side represents only disconnected diagrams, while the
right hand side represents only connected diagrams.\vspace{-0.4cm}

\begin{figure}[H]
  \begin{center}
    \begin{tabular}{c|c|c|c}
      \resizebox{0.2\columnwidth}{!}{\includegraphics{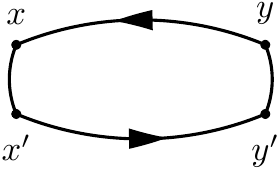}}
      &
      \resizebox{0.2\columnwidth}{!}{\includegraphics{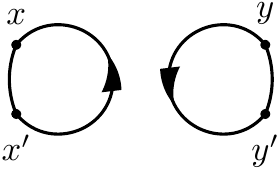}}
      &
      \resizebox{0.2\columnwidth}{!}{\includegraphics{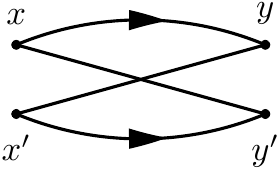}}
      &
      \resizebox{0.2\columnwidth}{!}{\includegraphics{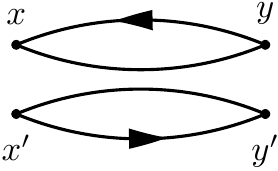}}
    \end{tabular}
    
\vspace{-0.7cm}
  \end{center}
  \caption{\label{fig:hlbl-con-disc}(First) Connected diagram $C (x, x', y,
  y') \sim e^{- m_{\pi} | x - y |}$. (Second) Disconnected diagram $D (x, x',
  y, y') \sim e^{- m_{\pi} | x - y |}$. (Third) Suppressed connected diagram
  $\mathcal{O} (e^{- 2 m_{\pi} | x - y |})$. (Fourth) Suppressed disconnected
  diagram $\mathcal{O} (e^{- 2 m_{\pi} | x - y |})$. The polarization indexes
  are always associated with the coordinates, and are thus omitted.}
\end{figure}

\vspace{-0.4cm}

In the hadronic light-by-light process, we have a hadronic four point
function. Let's denote the four points to be $x$, $x'$, $y$, $y'$. In the
region where $x$ is far separated from $y$, but $x'$ is close to $x$ and $y'$
is close to $y$, we know that the four point function should be dominated by
the $\pi^0$ exchange process. In order to deal with this hadronic four point
function, we first consider a four-point function in terms of the non-singlet
current operator $V_{{\mu}} (x) = \bar{u} \gamma_{{\mu}} u (x) -
\bar{d} \gamma_{{\mu}} d (x)$. It can be expressed in terms of the
diagrams shown in Figure \ref{fig:hlbl-con-disc}.\vspace{-0.1cm}
\begin{eqnarray}
  \langle V_{{\mu}} (x) V_{\nu} (x') V_{\rho} (y) V_{\sigma} (y') \rangle
  & = & 2 [C (x, x', y, y') + C (x, x', y', y) + C (x', x, y, y') + C (x', x,
  y', y)] \nonumber\\
  &  & \hspace{1cm} + 4 D (x, x', y, y') +\mathcal{O} (e^{- 2 m_{\pi} | x - y
  |}) .  \label{eq:v4-expansion}
\end{eqnarray}
Since the operator $V$ has isospin $I = 1$, the state created by the product
$V_{{\mu}} (x) V_{\nu} (x')$ or $V_{\rho} (y) V_{\sigma} (y')$ has isospin
$I = 0$ or $I = 2$. Thus, the energy of this state is at least $2 m_{\pi}$,
and the four point function must behave like $\mathcal{O} (e^{- 2 m_{\pi} | x
- y |})$. When combined with Eq. (\ref{eq:v4-expansion}), we learn
that:\vspace{-0.1cm}
\begin{eqnarray}
  D (x, x', y, y') & = & - \frac{1}{2}  [C (x, x', y, y') + C (x, x', y', y) +
  C (x', x, y, y') + C (x', x, y', y)] \nonumber\\
  &  & \hspace{3cm} +\mathcal{O} (e^{- 2 m_{\pi} | x - y |}) . 
  \label{eq:hlbl-disc-con}
\end{eqnarray}
We can see that this is very similar to Eq. (\ref{eq:neutral-pion}). This is
because the product of the two nearby current operators $[\bar{u}
\gamma_{{\mu}} u (x)] [\bar{u} \gamma_{{\mu}} u (x')]$ can be viewed
as a non-local version of $\bar{u} \gamma_5 u (x)$. Now, we can come back to
our original four-point function with $J_{{\mu}} = e_u  \bar{u}
\gamma_{{\mu}} u (x) + e_d  \bar{d} \gamma_{{\mu}} d
(x)$.\vspace{-0.1cm}
\begin{eqnarray}
  &  & \langle J_{{\mu}} (x) J_{\nu} (x') J_{\rho} (y) J_{\sigma} (y')
  \rangle  \label{eq:hlbl-neutral-pion}\\
  &  & \hspace{1cm} = (e_u^4 + e_d^4)  [C (x, x', y, y') + C (x, x', y', y) +
  C (x', x, y, y') + C (x', x, y', y)] \nonumber\\
  &  & \hspace{2cm} + (e_u^2 + e_d^2)^2 D (x, x', y, y') +\mathcal{O} (e^{- 2
  m_{\pi} | x - y |}) + \text{sub-leading disconnected diagrams} . \nonumber
\end{eqnarray}
\vspace{-0.8cm}

\begin{figure}[H]
  \begin{center}
    \begin{tabular}{c|c|c|c}
      \resizebox{0.2\columnwidth}{!}{\includegraphics{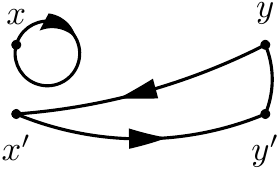}}
      &
      \resizebox{0.2\columnwidth}{!}{\includegraphics{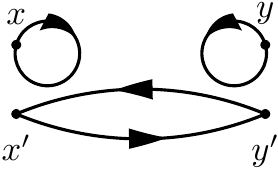}}
      &
      \resizebox{0.2\columnwidth}{!}{\includegraphics{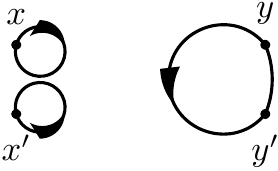}}
      &
      \resizebox{0.2\columnwidth}{!}{\includegraphics{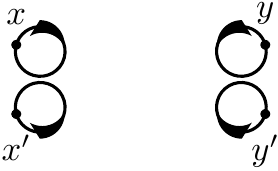}}
    \end{tabular}
    
\vspace{-0.7cm}
  \end{center}
  \caption{\label{fig:hlbl-sub-disc}Various kinds of sub-leading disconnected
  diagrams.}
\end{figure}

\vspace{-0.4cm}

Here the sub-leading disconnected diagrams refer to the diagrams vanishing in
$\ensuremath{\operatorname{SU}}(3)$ limit, some of them are shown in Figure
\ref{fig:hlbl-sub-disc}. These diagrams are only responsible for a small
portion of the pion exchange contribution, since they only couple to $\pi^0$
through the disconnected diagrams in the $\pi^0 \rightarrow \gamma \gamma$
process, and the disconnected diagrams in this process are known to be very
small {\cite{Feng:2012ck,Gerardin:2016cqj}}. Nevertheless,
contributions from these diagrams do not affect the ratio between the leading
disconnected diagrams and the connected diagrams. Combining Eq.
(\ref{eq:hlbl-disc-con})(\ref{eq:hlbl-neutral-pion}), we obtained that:
\begin{eqnarray}
  \frac{\ensuremath{\operatorname{dHLbL}} (x, x', y,
  y')}{\ensuremath{\operatorname{cHLbL}} (x, x', y, y')} & = & \frac{(e_u^2 +
  e_d^2)^2 \left( - \frac{1}{2} \right)}{(e_u^4 + e_d^4)} +\mathcal{O} (e^{-
  m_{\pi} | x - y |}) = \frac{- 25}{34} +\mathcal{O} (e^{- m_{\pi} | x - y |})
  . 
\end{eqnarray}
In comparison, {\cite{Bijnens:2016hgx}}, by assuming $\pi^0$ exchange
dominance and investigating the large $N_c$ behavior, independently estimated
that the total disconnected contribution to HLbL is large and negative with a
ratio of disconnected to connected of $- 25 / 34$.\vspace{-0.1cm}

A similar relation also appears in the hadronic vacuum polarization (HVP)
calculation, which was already known in Ref {\cite{Francis:2013qna}}. since
the lightest $I, J = 0, 1$ state has three pions, while the lightest $I, J =
1, 1$ state only has two pions, we know that for far separated $x$ and $y$, we
have
\begin{eqnarray}
  \langle (\bar{u} \gamma_{{\mu}} u - \bar{d} \gamma_{{\mu}} d) (x) 
  (\bar{u} \gamma_{{\mu}} u - \bar{d} \gamma_{{\mu}} d) (y) \rangle &
  \sim & e^{- 2 m_{\pi} | x - y |}, \\
  \langle (\bar{u} \gamma_{{\mu}} u + \bar{d} \gamma_{{\mu}} d) (x) 
  (\bar{u} \gamma_{{\mu}} u + \bar{d} \gamma_{{\mu}} d) (y) \rangle &
  \sim & e^{- 3 m_{\pi} | x - y |} . 
\end{eqnarray}
Following the same logic, when $x$ and $y$ are far separated:
\begin{eqnarray}
  \langle \bar{u} \gamma_{{\mu}} u (x) \bar{d} \gamma_{{\mu}} d (y)
  \rangle & \approx & - \frac{1}{2} \langle \bar{u} \gamma_{{\mu}} u (x)
  (\bar{u} \gamma_{{\mu}} u - \bar{d} \gamma_{{\mu}} d) (y) \rangle [1
  +\mathcal{O} (e^{- m_{\pi} | x - y |})] . 
\end{eqnarray}
With this relation, we know that the ratio between the disconnected HVP
contribution and the connected HVP contribution at large $x$, $y$ separation
is:
\begin{eqnarray}
  \frac{\ensuremath{\operatorname{dHVP}} (x,
  y)}{\ensuremath{\operatorname{cHVP}} (x, y)} & = & \frac{(e_u + e_d)^2
  \langle \bar{u} \gamma_{{\mu}} u (x) \bar{d} \gamma_{{\mu}} d (y)
  \rangle}{(e_u^2 + e_d^2) \langle \bar{u} \gamma_{{\mu}} u (x) (\bar{u}
  \gamma_{{\mu}} u - \bar{d} \gamma_{{\mu}} d) (y) \rangle} = \frac{-
  1}{10} +\mathcal{O} (e^{- m_{\pi} | x - y |}) . 
\end{eqnarray}
\vspace{-0.8cm}\subsection{Charged pion loop contribution}\vspace{-0.2cm}

Although it is estimated that the charged pion loop does not play a
significant role in the HLbL process {\cite{Jegerlehner:2009ry}}, its
contribution can be unambiguously identified when all four points in the
hadronic four-point-function are far separated.

There are three types of diagrams which contribute through the charged pion
loop : 1) the connected diagram, 2) the leading disconnected diagram, 3) the
sub-leading disconnected diagrams with two quark loops, where one loop is
attached to only one photon, the other loop is attached to three photons. We
denote their contribution to HLbL without the charge factors by $C$, $D$,
$D'$. The overall contribution is:\vspace{-0.4cm}
\begin{eqnarray}
  (e_u^4 + e_d^4) C + (e_u^2 + e_d^2)^2 D + (e_u + e_d)  (e_u^3 + e_d^3) D' &
  \propto & (e_u - e_d)^4 . 
\end{eqnarray}
Since we study the region where the charged pion loop dominates, we expect the
sum of all diagrams is proportion to the pion charge to the forth power,
because there are four photon - charged pion vertices. Above relations should
be true for any value of $e_u$ and $e_d$. According to this, we can derive
that $D \approx \frac{3}{2} C$, and $D' \approx - 2 C$. As a result, in the
region where all four points of the four-point function are far separated, the
ratio between the leading disconnected diagram and the connected diagram
is:\vspace{-0.4cm}
\begin{eqnarray}
  \frac{\ensuremath{\operatorname{dHVP}}}{\ensuremath{\operatorname{cHVP}}} &
  = & \frac{(e_u^2 + e_d^2)^2 D}{(e_u^4 + e_d^4) C} \approx \frac{- 75}{- 34}
  . 
\end{eqnarray}
One can easily apply this method to the problems discussed in the previous
subsection and obtain exactly the same result. However, this arguement,
relying on the assumption that the coupling is proportion to hadron charge, is
not as rigorous as the method discussed above, and one cannot easily estimate
the size of remainly effects like we did in the previous
subsection.\vspace{-0.6cm}

\section{Conclusions and ackownlegements}\vspace{-0.3cm}

We briefly summarized the lattice calculation of the hadronic light-by-light
contribution to muon anomalous magnetic moment, including the leading
disconnected diagram. The calculation follows the method developed in previous
work {\cite{Jin:2015eua,Blum:2015gfa}}, and this calculation is also
described in Ref {\cite{Blum:2016lnc}}. We also discussed the large separation
behaviour of the four-point function within the hadronic light-by-light
diagram, and obtained the theoretical ratio between the connected and the
leading disconnected diagrams in different situations. We plan to a) address
the discretization errors by computing on our finer, physical-pion-mass $64^3$
lattice with similar physical volume. b) address the finite volume effect by
using the $48^3$ QCD ensemble inside a larger QED box or infinite volume
analytic formula. and c) compute additional sub-leading disconnected diagrams.

We would like to thank our RBC and UKQCD collaborators for helpful discussions
and support. We would also like to thank RBRC and BNL for BG/Q computer time.
The $48^3$ computation is performed on Mira with ALCC allocation using BAGEL
{\cite{Boyle:2009vp}} library. The CPS {\cite{Jung:2014ata}} software package
is also used in the calculation. The computation is performed under the ALCC
Program of the US DOE on the Blue Gene/Q (BG/Q) Mira computer at the Argonne
Leadership Class Facility, a DOE Office of Science Facility supported under
Contract De-AC02-06CH11357. T.B is supported by U.S. DOE grant
\#DE-FG02-92ER41989. N.H.C and L.C.J are supported by U.S. DOE grant
\#DE-SC0011941. M.H is supported by Grants-in-Aid for Scientific Research
\#25610053. T.I, C.J. and C.L are supported by U.S. DOE Contract
\#AC-02-98CH10996(BNL). T.I. is supported in part by the Japanese Ministry of
Education Grant-in-Aid, No. 26400261. CL acknowledges support through a DOE
Office of Science Early Career Award.